# Title: Tailoring interfacial nanostructures at graphene/transition metal dichalcogenide heterostructures

**Authors:** Ya-Ning Ren[§], Mo-Han Zhang[§], Qi Zheng[§], Lin He[†]

**Affiliations:**

Center for Advanced Quantum Studies, Department of Physics, Beijing Normal University, Beijing, 100875, People's Republic of China

[§]These authors contributed equally to this work.

[†]Correspondence and requests for materials should be addressed to Lin He (e-mail: helin@bnu.edu.cn).

**Integration of two-dimensional (2D) van der Waals (vdWs) materials with non-2D materials to realize mixed-dimensional heterostructures has potential for creating functional devices beyond the reach of existing materials and has long been a pursuit of the material science community. Here we report the patterning of monolayer/bilayer transition metal dichalcogenide (TMD) nanostructures with nanometer-precision, tunable sizes and sites at interfaces of graphene/TMD heterostructures. Our experiments demonstrate that the TMD nanostructures can be created at selected positions of the interface using scanning tunneling microscope lithography and, more importantly, the sizes of the interfacial TMD nanostructures can be tuned with nanoscale precise by merging adjacent TMD nanostructures or by further tailoring. A structural phase transition from hexagonal phase in bulk TMD to monoclinic phase in the interfacial nanostructures is explicitly observed, which locally introduce well-defined electrostatic potentials on graphene. These results not only enable the creation of high-quality patterned p-n junctions in vdWs heterostructures,**

**but also provide a new route to realize custom-designed mixed-dimensional heterostructures.**

Local control of doping type and carrier concentration are essential requirements to realize nanoscale electronic and optoelectronic devices in two-dimensional (2D) van der Waals (vdWs) heterostructures[1-7]. They are also vital to study the fundamental physics[8-25], such as the Klein tunneling and atomic collapse. In spite of great efforts on the related study during the past ten years, locally patterning high-quality p-n junctions with well-defined sizes and sites in 2D vdWs heterostructures remain outstanding challenges in experiment up to now. Here, we overcome the challenges by combining the ability to build high-quality artificial vdWs heterostructures with the well-developed scanning tunneling microscope (STM) lithography[26-28]. By using the STM lithography, we create monolayer/bilayer transition metal dichalcogenide (TMD) nanostructures with tunable sizes and sites at interfaces of graphene/TMD heterostructures. The interfacial TMD nanostructures introduce well-defined electrostatic potentials on graphene, allowing us to fully engineer electronic structure of the heterostructures. Moreover, the method reported here helps us to realize a wide variety of high-quality mixed-dimensional heterostructures[29,30] by integration of the 2D graphene/TMD heterostructures with the TMD nanostructures.

In our experiment, four different high-quality graphene/TMD heterostructures are obtained by using transfer technology, as reported previously[22,31,32], of graphene monolayer onto mechanical-exfoliated TMD sheets, $WSe_2$, $WS_2$, $MoSe_2$, and $MoTe_2$. Besides the capability of atomic-resolution imaging, the STM has demonstrated the ability to modify surface of the studied systems locally[26,27]. Here we show that it is possible to locally modify interfacial structures of the graphene/TMD heterostructures through voltage pulses applied to STM tip, as schematically shown in Figs. 1b-1c. By using STM tip pulses, the supporting TMD substrate can be locally fractured and nanoscale pits are generated in the substrates. Then we are able to "create" monolayer/bilayer TMD nanostructures at the interface from the pits in the substrates. As an example, Fig. 1d summarizes the fracture probability of the $MoSe_2$ substrate in the graphene/$MoSe_2$ heterostructure with different experimental parameters. Generally, the fracture probability is higher for larger tip pulse. In our experiment, nanoscale pits of the bulk TMD substrates are usually observed around the interfacial TMD nanostructures (See Figs. S1-S3 in the supplementary information), suggesting that the interfacial TMD nanostructures are generated from the pits of the substrate by the tip pulses. Figure 1e shows a representative STM image of a monolayer $WSe_2$ nanostructure generated at the interface of graphene/$WSe_2$ heterostructure and the thickness of the nanostructure is the same as that of a $WSe_2$ monolayer $\sim 0.8$ nm. The topmost graphene monolayer is un-attacked by the tip pulses and there is no defect in the graphene and no graphene fragment around the $WSe_2$ pits and nanostructures. Such a result is quite reasonable because that the C-C bond in graphene is rather strong and the fracture stress of graphene is about 6 times larger than that of $WSe_2$[33].

The $WSe_2$ nanostructures can introduce Coulomb-like electrostatic potentials on the graphene above them, which generate quasibound states of massless Dirac fermions in graphene through both the whispering gallery modes (WGMs) and the atomic collapse states (ACSs)[22]. Figure 1f shows a representative scanning tunneling spectroscopy (STS), *i.e.*, d$I$/d$V$, spectroscopic map along the red dashed line in Fig. 1e, and Fig. 1g shows two typical STS spectra measured in the graphene on and off the $WSe_2$ nanostructures. In graphene off the $WSe_2$ nanostructures, a typical V-shaped spectrum can be observed. Whereas, a sequence of temporarily-confined quasibound states, shown as resonance peaks in the spectrum, are clearly observed in graphene on the $WSe_2$ nanostructures. The ACSs locate at the center of the potential field due to collapse by the Coulomb potential, while the WGMs locate at the edge of the potential profile, forming concentric ring structures[22], as shown in typical STS maps in Fig. 1h. Therefore, the $WSe_2$ nanostructures introduce well-defined electrostatic potentials in the heterostructures. Although the quasibound states in graphene/$WSe_2$ heterostructures are well studied, two central issues: (1) how to controllably create the interfacial $WSe_2$ nanostructures and (2) the origin of the electronic potential generated by the interfacial $WSe_2$ nanostructures in the graphene/$WSe_2$ heterostructure, are not clear[22]. Here the first key issue is well addressed and we demonstrate the ability to create the interfacial $WSe_2$ nanostructures with controllable sites in the heterostructure. Moreover, by pre-setting the coordinates of the STM tip and using a series of voltage pulses, the positions of the interfacial $WSe_2$ nanostructures can be created with nanoscale precision in their sites and different custom-designed patterns of the nanostructures can be obtained, as shown in Fig. 1i as an example. Here we should point out that similar mixed-dimensional heterostructures can be obtained in all the studied graphene/TMD heterostructures by using voltage pulses applied to STM tip (see Figs. S1-S4). Therefore, our experiments demonstrate a general route to realize a wide variety of mixed-dimensional heterostructures by integration of the 2D graphene/TMD heterostructures with the monolayer/bilayer TMD nanostructures.

The second unexplored issue is why the interfacial TMD nanostructures can locally change the doping of graphene. To answer this question, we explore atomic structures of hundreds of the interfacial TMD nanostructures and our experiments demonstrate that the local electrostatic potential is closely related to the phase transition of the TMD nanostructures. Here we still introduce the result of graphene/$WSe_2$ heterostructure as an example and the results obtained in other graphene/TMD heterostructures are similar. In our experiment, the structure of the $WSe_2$ substrate is characterized by a trigonal-prismatic coordination of transition metal atoms and it is denoted as the 1H (2H) phase in the single-layer (bulk) form (Fig. 2a). However, our atomic resolution STM measurements indicate that only several percent of all the studied interfacial $WSe_2$ nanostructures remains in the 1H phase and the rest of the interfacial $WSe_2$ nanostructures becomes monoclinic 1T' phase after the tip pulse. There are many possible reasons, such as local electrostatic gating induced by the STM tip, local electron doping and mechanical strain induced by the adjacent graphene and $WSe_2$ substrate, for the observed structural phase transition in the interfacial $WSe_2$ nanostructures[34]. Although the $WSe_2$ nanostructure is covered by a graphene sheet, it is still possible to obtain its atomic-resolved STM image. Figure 2b shows a representative result of the 1H-phase $WSe_2$ nanostructure. For the graphene covering the 1H-phase $WSe_2$ nanostructure, both the supporting substrates of graphene on and off the nanostructure are the hexagonal-phase $WSe_2$ and, in this case, we only observe a very small variation of the local doping, which may arise from effects of dangling bonds at the edge of the nanostructures. Then, it is difficult to observe a series of quasibound states in graphene above the 1H-phase WSe$_2$ nanostructure (See Figs. S5-S6). For the graphene covering the 1T'-phase WSe$_2$ nanostructure, the supporting substrates of graphene on and off the nanostructure are the 1T'-phase and 2H-phase WSe$_2$ respectively, which have quite different electronic properties and work functions[34]. Then, we observe large variations of the local doping in graphene on and off the WSe$_2$ nanostructures (See Figs. 1e-1g and Figs. S5-S6). Figure 2c (left panel) shows a typical atomic-resolved STM image around a 1T'-phase WSe$_2$ nanostructure of the heterostructure. The nanostructure is in a single domain of the 1T'-phase phase with a uniform orientation of atomic rows (See Fig. S7 for more experimental data). Because of the C$_3$-symmetric 1T-phase parent phase, the 1T'-phase WSe$_2$ can be formed in three equivalent directions (Fig. 2d) and is predicted to be a ferroelastic material[35]. In our experiment, the single-domain nanostructure can be reversibly switched to a nanostructure with multiple 120° domain boundaries, as shown in Fig. 2c (right panel) and Fig. S8, through the application of a small bias scan by the STM (see Figs. S9-10 for more experimental data). The STM measurements will remove adsorbate clusters of the interfacial nanostructures or change the interaction between the interfacial nanostructures and graphene, which may change the local strain distribution and result in the transition between multiple domains and the single domain. Such a phenomenon provides evidence of the ferroelastic nature of the 1T'-phase WSe$_2$ nanostructure[36].

In previous studies, the STM has been demonstrated the ability to manipulate atoms[19], molecules[37], and even nanostructures[38,39] at the surface of the samples. Although the WSe$_2$ nanostructures are buried at the interface of the graphene/WSe$_2$ heterostructures, here we show that they still can be easily manipulated with an STM tip, as summarized in Fig. 3. Figures 3a-3c show schematics of the process of manipulation. By reducing the distance between the STM tip and the graphene, we can lift the graphene beneath the tip. The WSe$_2$ nanostructure will also be lifted because the vdWs interaction between graphene and the nanostructure (Fig. 3b). By moving the tip along a given direction and then increasing the tip-graphene distance, the WSe$_2$ nanostructure can be moved to a selected position (Fig. 3c). Figure 3d summarizes a representative result obtained in our experiment and we demonstrate the ability to tune positions of the interfacial WSe$_2$ nanostructures with nanoscale precise (See Fig. S11 for more experimental results). Because of the small fracture stress of the WSe$_2$[33], we further show the ability to tailor the interfacial nanostructures by using the STM tip. Figure 3e shows a typical result obtained in our experiment. First, the selected WSe$_2$ nanostructure is moved to a nearby pit of the WSe$_2$ substrate. The edge atoms of the pit seem to increase the interaction between the WSe$_2$ nanostructure and the WSe$_2$ substrate. Then, when the WSe$_2$ nanostructure is lifted by the STM tip through vdWs interaction between graphene and the nanostructure, it can be tailored into two pieces. As shown in Fig. 3e, the WSe$_2$ nanostructure is tailored into two pieces by lifting the STM tip and one piece is pinned at the pit of the WSe$_2$ substrate (See Fig. S11 for more experimental results). Combining the ability to tune the sites and the ability to tailor the sizes of the interfacial nanostructures, the method reported here may open a new avenue to build custom-designed patterns of the WSe$_2$ nanostructures to realize artificial model systems via coupled quasibound states in the graphene.

An obvious result obtained in our experiment is that the area of the interfacial $WSe_2$ nanostructure is larger than that of the $WSe_2$ pit, as shown in Figs. 1i, 3e, and Figs. S1-S3. Figure 4a summarizes the area of 20 $WSe_2$ nanostructures and their corresponding nearby $WSe_2$ pits obtained in our experiment. Generally, the area of the interfacial $WSe_2$ nanostructure is several times larger than that of the corresponding nearby $WSe_2$ pit. Because that the interfacial $WSe_2$ nanostructure is generated from the pit of the $WSe_2$ substrate, two important results can be deduced from such an experimental result. First, the depth of the $WSe_2$ pits should be much larger than the thickness of monolayer $WSe_2$ (although the pits is covered by a topmost graphene sheet, the depth of some pits is measured to be larger than the thickness of monolayer TMD, as shown in Figs. S1.). Second, the larger interfacial $WSe_2$ nanostructure should be generated by merging the smaller $WSe_2$ fragments from the pit after the tip pulse. With considering the single-crystal nature of the interfacial $WSe_2$ nanostructure, such a result implies that there is interfacial "chemical reaction" between the small $WSe_2$ fragments to ensure that they merge into a larger $WSe_2$ nanostructure after the tip pulse. To confirm this, we further apply several tip pulses around an interfacial $WSe_2$ nanostructure, as shown in Fig. 4b. With increasing the number of tip pulses, the area of the adjacent $WSe_2$ nanostructure increases. Such a result demonstrates explicitly that there is interfacial "chemical reaction" between the small $WSe_2$ fragments at the interface and they will merge into a larger single-crystal $WSe_2$ nanostructure.

In summary, we report the patterning of TMD nanostructures with nanometer-precision, tunable sizes and sites at interfaces of graphene/TMD heterostructures by using STM lithography. The reported method helps us to realize a wide variety of high-quality mixed-dimensional heterostructures by integration of the 2D graphene/TMD heterostructures with the TMD nanostructures, which opens a new avenue to fully engineer electronic structure of the heterostructures.

**Acknowledgments:**

This work was supported by the National Key R and D Program of China (Grant Nos. 2021YFA1401900, 2021YFA1400100) and National Natural Science Foundation of China (Grant Nos. 12141401,11974050).

**Author contributions**

Y.N.R., M.H.Z. and Q.Z. performed the sample synthesis, characterization and STM/STS measurements. Y.N.R. and L.H. analyzed the data. L.H. conceived and provided advice on the experiment and analysis. Y.N.R. and L.H. wrote the paper with the input from others. All authors participated in the data discussion.

**Data availability statement**

All data supporting the findings of this study are available from the corresponding author upon request.

**Methods**

**CVD Growth of Graphene.** The large area graphene monolayer films were synthesized on a 20 × 20 mm$^2$ polycrystalline copper (Cu) foil (Alfa Aesar, 25 μm thick) via a low-pressure chemical vapor deposition (LPCVD) method. Ar flow of 50 sccm (Standard Cubic Centimeter per Minutes) and H$_2$ flow of 50 sccm were maintained throughout the whole growth process. The Cu foil was heated from room temperature to 1030 ℃ in 30 min and annealed at 1030 ℃ for 12 hours. Then CH$_4$ flow of 5 sccm was introduced for 20 min to grow high-quality large area graphene monolayer. Finally, the sample was cooled down naturally to room temperature.

**Construction of graphene/TMD heterostructure.** The as-grown graphene film was transferred onto TMD substrates by the polymethyl methacrylate (PMMA)-assisted method. PMMA was first uniformly coated on Cu foil with graphene monolayer. Then, the sample was immersed into the ammonium persulfate solution to etch the underlying Cu away, so the PMMA/graphene film was detached from the Cu foil. The PMMA/graphene film was cleaned in deionized water for hours. The bulk TMDs (Shanghai Onway Technology Co., Ltd) were separated into thick-layer TMD sheets by traditional mechanical exfoliation technology and then transferred to SiO$_2$/Si wafer. We placed PMMA/graphene onto SiO$_2$/Si wafer which has been transferred with TMD sheets in advance. Finally, the PMMA was removed by acetone and then annealed in low pressure with Ar flow of 50 sccm and $H_2$ flow of 50 sccm at ~300 °C for 1 hours.

**STM and STS Measurements.** STM/STS measurements were performed in low-temperature (77 K and 4.2 K) and ultrahigh-vacuum (~$10^{-10}$ Torr) scanning probe microscopes [USM-1500 (77 K) and USM-1300 (4.2 K)] from UNISOKU. The tips were obtained by chemical etching from a tungsten wire to minimize tip-induced band bending effects of graphene. The differential conductance ($\mathrm{d}I/\mathrm{d}V$) measurements were taken by a standard lock-in technique with an ac bias modulation of 5 mV and 793 Hz signal added to the tunneling bias.

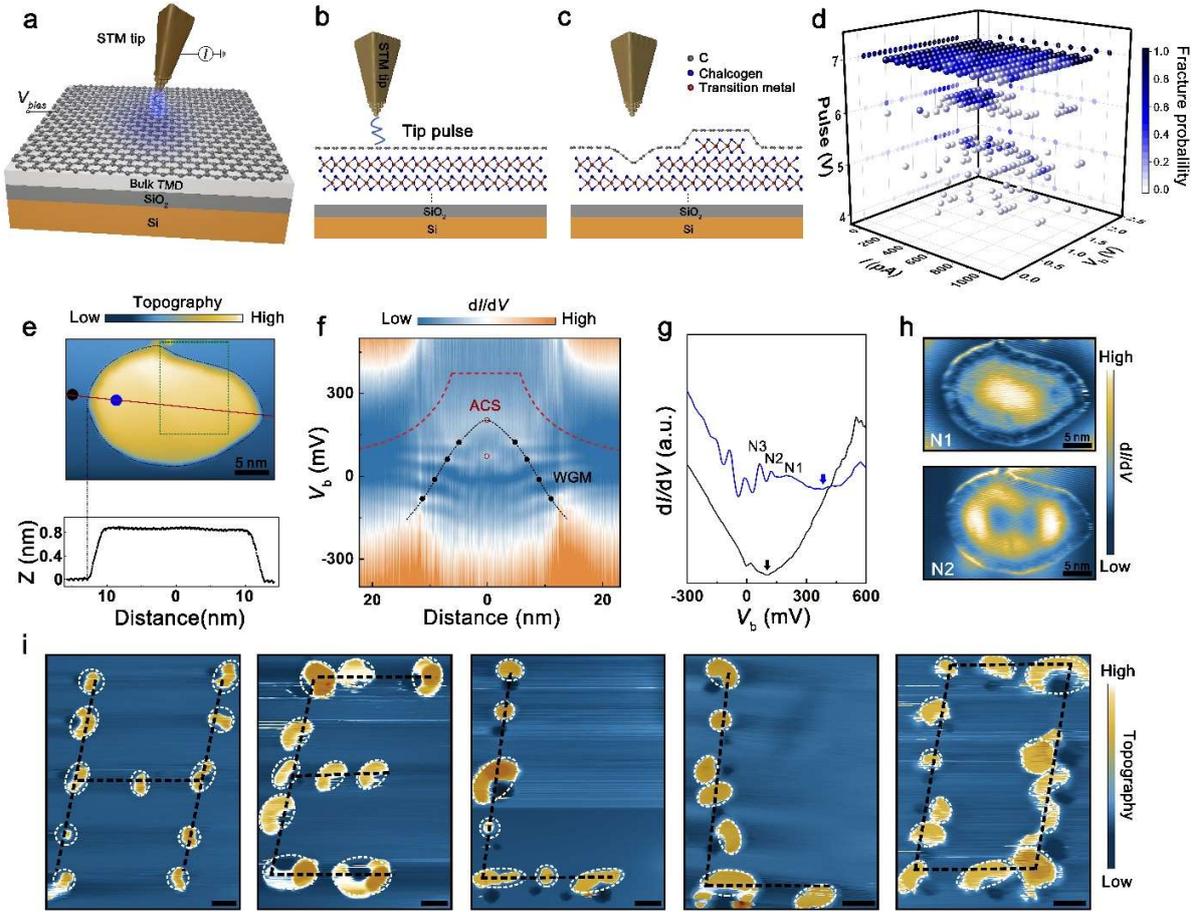

**Fig. 1. In situ creating patterns of interfacial TMD nanostructures in graphene/TMD heterostructures. a.** Sketch of the STM setup. (**b**) and (**c**), Schematics of graphene/TMD heterostructures before and after an applied tip pulse, respectively. The STM tip pulse creates an interfacial TMD nanostructure accompanied by a nanoscale pit in the substrate. **d.** The fracture probability of TMD substrates (here, the result is obtained in graphene/MoSe$_2$ heterostructure as an example) with different parameters of the STM tip pulse. Each point in this figure is obtained by counting fracture probability of 20 tip pulses and the total number of pulses is 1400 to obtain the result. **e.** A STM image ($V_{sample}$ = 350 mV, $I$ = 400 pA) of an interfacial WSe$_2$ nanostructure. Bottom: the height profile shows that the height of the WSe$_2$ nanostructure is ~ 800 pm. **f.** The d$I$/d$V$ spectroscopic map versus the spatial position along the direction of red solid line in panel e. The red dashed line indicates Dirac point energy. The black solid dots indicate the quasibound

states via the WGMs confinement, and the red hollow dots indicate the ACSs. **g**. Typical d$I$/d$V$ spectra measured in graphene on (blue line) and off (black line) the interfacial WSe$_2$ nanostructure marked in panel e. The arrows denote the positions of the Dirac point. **h**. STS maps recorded at different energies (N1 and N2 marked in panel g) of the interfacial WSe$_2$ nanostructure. **i**. Patterns of interfacial WSe$_2$ nanostructures created by a series of STM tip pulses. The scale bar is 30 nm in panel i.

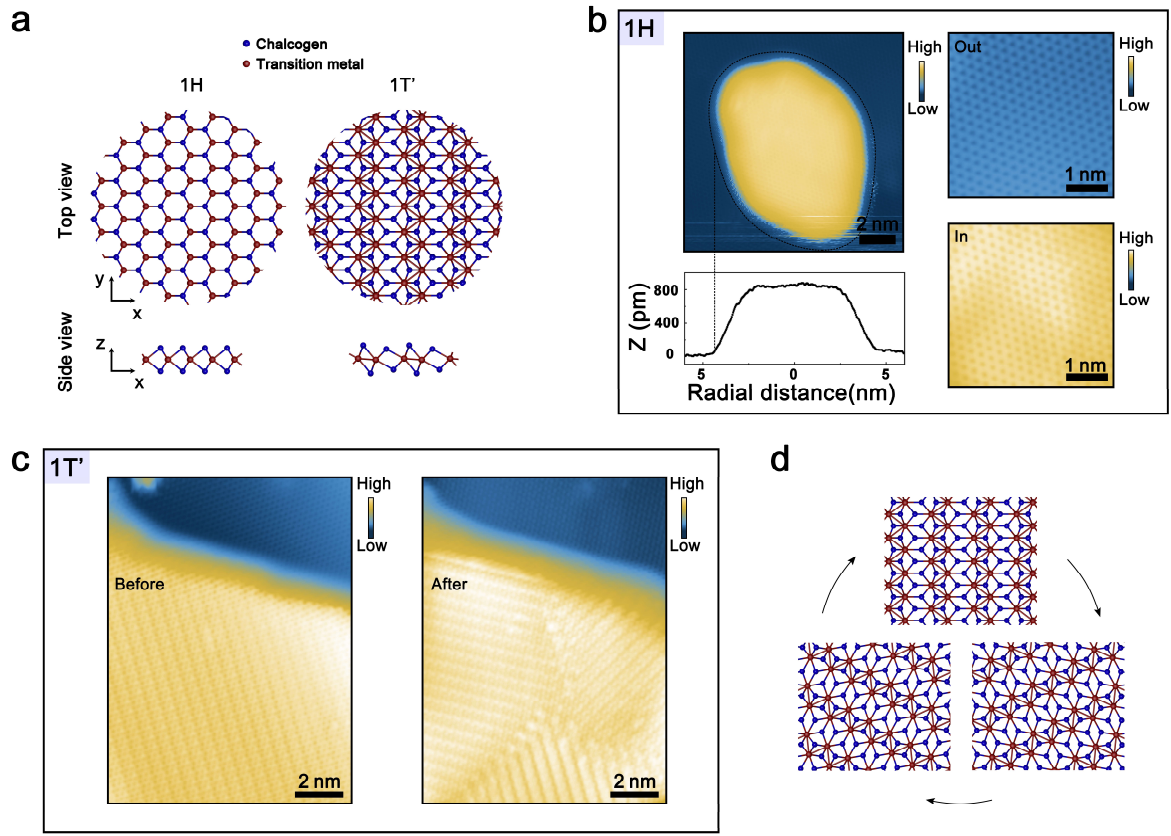

**Fig. 2. Structures of interfacial 1H-phase and 1T'-phase WSe₂ nanostructures. a.** Atomistic structures of monolayer TMDs in their trigonal prismatic (1H) and monoclinic (1T') phases. In the 1T' phase, the distorted M atoms form 1D zigzag chains. **b.** A representative result of the graphene covering the 1H-phase WSe₂ nanostructure. Left panel: STM image ($V_{sample} = 300$ mV, $I = 300$ pA) and height profile of the graphene covering the 1H WSe₂ nanostructure. Right panel: Atomic-resolved STM image on (bottom panel) and off (top panel) the interfacial WSe₂ nanostructure. **c.** Left panel: A zoom-in STM image ($V_{sample} = 350$ mV, $I = 400$ pA) of green dashed square in Fig.1e, showing the parallel dimerization orientation of the 1T'-phase WSe₂. Right panel: The nanostructure with multiple 120° domain boundaries after a small bias scan by the STM. **d.** Schematic diagram of three different dimerization orientations of the 1T'-phase TMDs, which can be switched between different dimerization orientations by applying stress.

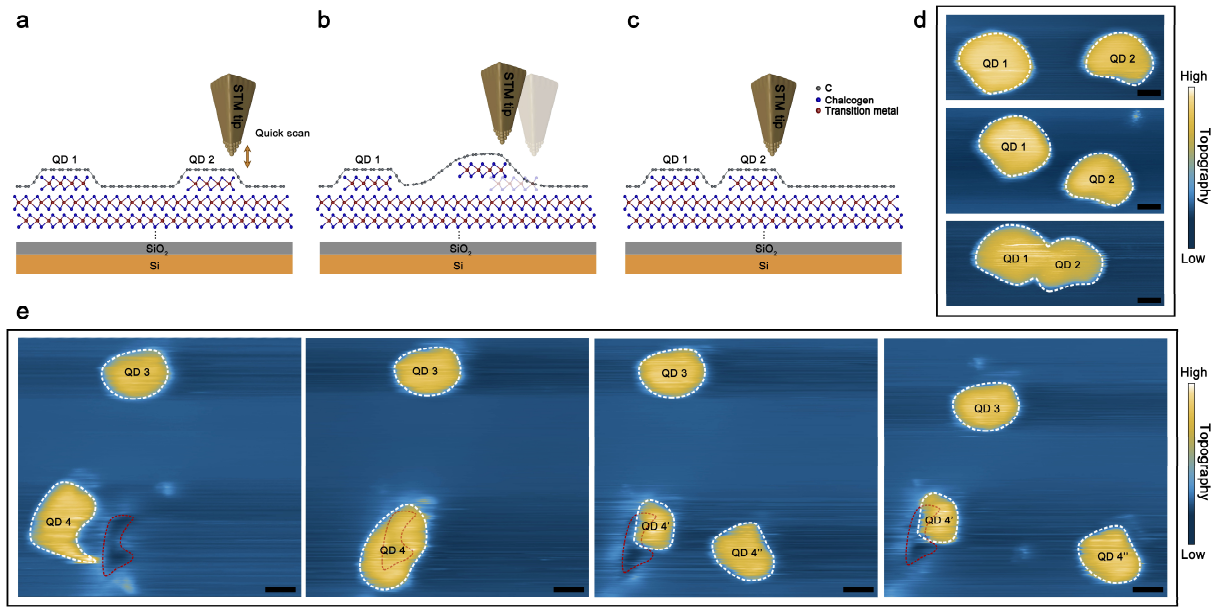

**Fig. 3. Manipulating and tailoring interfacial WSe₂ nanostructures. a-c.** Schematics of moving interfacial TMD nanostructures. By reducing the tip-graphene distance, graphene beneath the STM tip can be lifted. The WSe₂ nanostructure will also be lifted due to the vdWs interaction between graphene and the nanostructure. By moving the tip along a given direction and then increasing the tip-graphene distance, the WSe₂ nanostructure can be moved to a selected position. **d.** A representative result showing the manipulation of two interfacial WSe₂ nanostructures. From top panel to bottom panel, the two nanostructures (QD 1 and QD 2) are gradually approaching and merging under the control of STM tip. The scale bar is 5 nm. **e.** A representative result of tailoring the interfacial WSe₂ nanostructures. From left panel to right panel, the nanostructure QD 4 is first moved to a nearby pit of the WSe₂ substrate (outlined in red dashed line). The edge atoms of the pit increase the interaction between the WSe₂ nanostructure and the WSe₂ substrate. Then, the QD 4 can be tailored into two pieces, QD 4' and QD 4", when it is lifted by the STM tip. The nanostructures QD 3 and QD 4" can be further manipulated by using the STM tip. The scale bar is 10 nm.

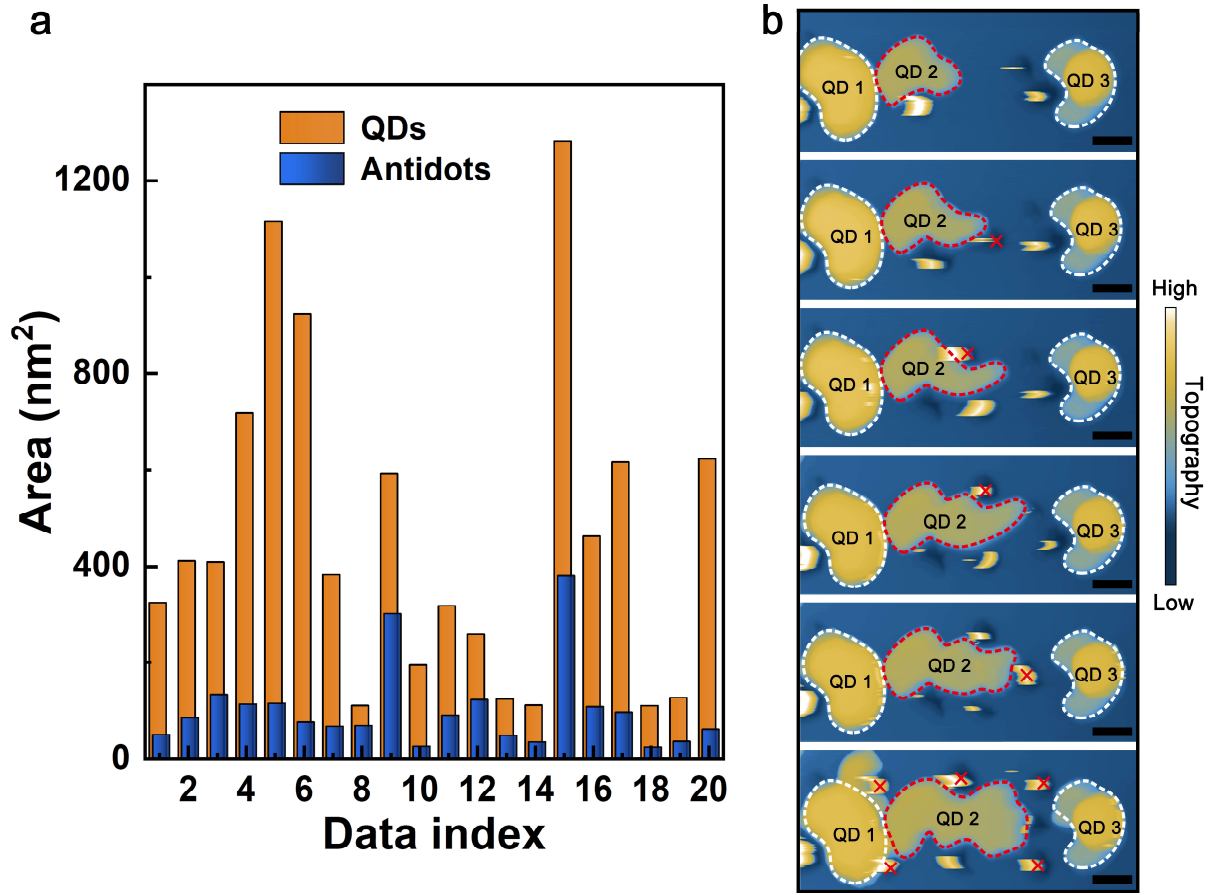

**Fig. 4. Merging the WSe₂ fragments into large nanostructures through interfacial "chemical reaction". a.** The area of WSe₂ nanostructures and their corresponding nearby WSe₂ pits. **b.** From top panel to bottom panel, the area of the nanostructure QD 3 increases with increasing the number of tip pulses (the positions are marked by "×"), demonstrating that the larger interfacial TMD nanostructures are merged from smaller fragments generated from the pits.